# Circuit architecture of a sub-GHz resolution panoramic C band on-chip spectral sensor

Mehedi Hasan,[1,*] Mohammad Rad,[2] Gazi Mahamud Hasan,[1] Peng Liu,[1] Patric Dumais,[2] Eric Bernier,[2] and Trevor Hall,[1]

[1]*Photonic Technology Laboratory, Centre for Research in Photonics, Advanced Research Complex, University of Ottawa, 25 Templeton Street, Ottawa, K1N 6N5, ON, Canada*
[2]*Huawei Technologies Canada, 303 Terry Fox Drive, Kanata, K2K 3J1, ON, Canada*
*\*mhasa067@uottawa.ca*

Monitoring the state of the optical network is a key enabler for programmability of network functions, protocols and efficient use of the spectrum. A particular challenge is to provide the SDN-EON controller with a panoramic view of the complete state of the optical spectrum. This paper describes the architecture for compact on-chip spectrometry targeting high resolution across the entire C-band to reliably and accurately measure the spectral profile of WDM signals in fixed and flex-grid architectures. An industry standard software tool is used to validate the performance of the spectrometer. The fabrication of the proposed design is found to be practical.

Software Defined Network (SDN), Elastic Optical Network (EON), and Artificial Intelligence (AI) have gained wide acceptance as they support network operators (Bell, Rogers, Verizon, etc.) in their quest to address the challenges of rapidly changing and demanding service requirements, while making efficient use of network resources [1,2]. SDN allows programmability of network functions and protocols. EON allow the allocation of an arbitrary and appropriate spectral range and modulation format to an optical path according to application bandwidth and quality of service requirements taking into account optical physical layer attributes such as impairments [3]. AI allows the network to provision resources in response to current service requests while learning from the past to improve network efficiency and effectiveness [4].

SDNs scale by control and adaptive management, and handle changing demand and resources to achieve energy, resource efficiency and sustainability. Applying SDN principles to the physical layer that includes optical components, such as wavelength switches, fibre, add/drop mux/de-mux, amplifiers, filters, and sensors; requires fully programmable functions be they implemented in optics or electronics. As a consequence of the rigidity of the physical layer infrastructure, SDN research has focused principally on the higher digital electronic layers of the network. Applying SDN to the optical transport network could enhance both operational management in terms of cost saving and service layer performance such as fast connection turn-up, margin squeezing or health monitoring of the network. The adaptability provided from AI combined with the flexibility of SDN & EON along with the extremely fast processing capability of optical devices will transform the existing transport networks into a next-generation SDN-enabled energy efficient optical transport network.

However, optical performance monitoring (OPM) is an important necessity in an optical network that can offer higher speeds, lower cost and reduced environmental impact. Network management agents or optimization algorithms require up to date telemetry data of the network on links, components and operating points of service (WDM traffic channels in case of transport optics). Monitoring information can be used for better resource optimization to maximize the reach versus rate. OPM information can also be used for performance prediction and planning in case of network reconfiguration, capacity scaling or network or component fault recovery. The monitoring information includes power, loss, bit-error-rate (BER), optical signal-to-noise-ratio (OSNR), electrical signal-to-noise ratio (ESNR), etc. In practice any parameter measured in the network can be used for the purpose of OPM. However, power is a strong indication of performance in optical systems as like other systems and hence OPM is used to refer to "power" in this manuscript.

Power measurement (monitoring) or spectrum sensing while seemingly very simple and basic is still challenging in transport optics especially in WDM networks. WDM channels typically are spread from 1530 nm to 1570 nm, i.e., around 40 nm of fiber optic bandwidth. Traditional WDM channels are spaced at known and fixed locations of the WDM spectrum (50 GHz or 100 GHz ITU spectrum). However, the current elastic optical networks are following more and more flex-grid channel profiles where the channel center wavelength can be any place in the spectrum with much finer resolutions (e.g. smaller than 6.25 GHz) and they also can have different power profiles (i.e., bandwidth and power spectral density). Power measurement is done by OPM cards traditionally where a tunable filter sweeps the spectrum (C band) typically with 50 GHz resolution and hence ITU channel power readings become available. OPM cards are large; consume considerable power; and, most important, are expensive. As a result, they are only available at few points in the network. Typically, reconfigurable add-drop modules (ROADMs) are equipped with OPM cards. Since existing power measurements are based on ITU, flex ready spectrum measurements are adopted for flex grid systems. New modules capable of performing power measurement for desired location (center wavelength) and desired resolution (known bandwidth) are being used in flex-grid ready ROADM architectures. They however still suffer from the cost issue and hence spectrum measurement is only available at add-drop nodes. ROADMs with high degrees usually share an OPM module for the purpose of spectrum measurement. This makes the OPM speed slow depending on the number of lines it supports. Amplifier nodes do not have power reading capabilities. Therefore, almost all analysis of photonic layer optimizations uses analytical, semi-analytical, or machine learned based modeling to estimate the performance of WDM channels in a section (ROADM-to-ROADM).

Reliable spectral measurement across the network is therefore a key enabling technology. Complete knowledge of the state of the network is a prerequisite to enabling SDN-EON-AI to make effective use of colour-, direction-, contention-, grid-less, filter-, gap-less ROADM, flexible channels centre frequencies and width, flexible sub-carriers in super-channels, flexible modulation formats and forward error control coding transponders, and impairment-aware wavelength routing and spectral assignment. The absence of OPM (with the focus of power measurement) makes it very difficult to have reliable performance estimation or have the models trained and fed with proper live and accurate measurement across the network for the purpose of performance optimizations. Knowledge of spectral content in a section of optical networks is of great value for operators and photonics infra-structure owners.

A large variety of on-chip spectrometer designs have been unveiled and realized over the past decade such as cascaded MRR-AWG architecture [5-6], cascaded MZI based processor [7], discrete Fourier transform infrared spectroscopy [8], and dispersive configurations (arrayed waveguide gratings (AWG) [9], echelle gratings (EG) [9] or cascaded micro-rings [10]. In [5] the authors propose a combined architecture where the ring and AWGs need to be tuned to match the ring response with the AWG center wavelength. Depending on the number of channels, fixed grid reading of the spectrum is achieved. Similar to the previous prior art the authors of [6] proposed a cascaded architecture where a ring resonator (RR) is followed by a parallel pair of arrayed waveguide gratings (AWG) with different center wavelengths to reduce the cross talk. The center wavelength of the AWGs must differ by half a channel spacing for proper operation but there is no mention of if and how the whole spectrum may be scanned. When the RR is not tuned to an AWG channel passband center, the spectrometer will suffer increased insertion loss and channel leakage (cross talk) with increased detuning. In [7] a fast-Fourier transformation (FFT)-like approach is used by performing multiple measurements of the filtered spectrum. The authors demonstrate a resolution of 23 pm (~ 2.9 GHz) over a 184 pm range about 1550 nm. Sub-GHz resolution seems to be within reach but entire C band coverage remains challenging as it requires multiple arms (up to 10) which imposes significant loss, complex control and measurement dynamics. Following a similar approach, ref. [8] proposes a new signal processing technique performed electro-optically where a combination of numerous measurements with data processing provides the desired high-resolution power measurement across a desired band. To achieve sub-GHz resolution more than 10 stages are needed. Therefore, this strategy and architecture suffers similarly in scaling when high resolution spectrum sensing over a wide band is needed. On the other hand, in [9] echelle gratings (EG) and AWGs have been proposed to construct spectrometers capable of operation over a broad range different wavelength in [9]. The resolution is restricted by the number of samples of the spectrum fixed number of EG/AWG channels A multiple ring resonators in cascade architecture has been proposed in [10] for high resolution spectrum measurement. These architectures have proven to be extremely complicated to control and calibrate in a reliable fashion due to the fact that two or more very sensitive components (RR) are needed to be tuned together. In summary, the existing measurement technologies fail to combine acceptable resolutions with wideband operation and do not support feasible integration with other products which limits their deployment mainly due to high cost, loss and foot print. In addition to that, the requirement of sub-GHz resolution bandwidth to reliably and accurately distinguish the individual carriers in a densely packed multi-carrier super-channels or to monitor the differential power of adjacent channels in fixed and flex-grid architectures with a large free-spectral range for transparent operation across the entire C band 1530 nm-1565 nm for an integrated solution remains challenging.

In this paper, the design of a technologically viable compact on-chip high resolution wideband spectrometer is presented and verified by software simulation using an industry standard tool. The target application is to measure the spectral profile of a WDM signals accurately in flex and fixed grid architectures across the entire C-band 1530 nm-1565 nm aiming at sub-GHz resolution bandwidth with minimum scan time and less than 1 GHz frequency accuracy. The photonic integration of the circuit architecture can be implemented using a mature fabrication technology; low index contrast Silica on Silica or CMOS compatible low loss $Si_3N_4$ photonic integration platform. However, to meet the technical specifications such as compact size and high-resolution bandwidth, the $Si_3N_4$ platform is preferred due to its impressive agreement between simulation and practical measurement for the components used in this design architecture [11-14].

The circuit architecture of the proposed spectrometer [15] is illustrated schematically in Fig. 1. It consists of three stages. The first stage is a ring resonator (RR) which has the function of defining the spectrometer resolution. It provides a periodic train of resonances each with bandwidth < $1 GHz$ and spaced by its free-spectral range (FSR). It is tuneable in frequency over one FSR. The final stage is an arrayed waveguide grating (AWG) which has the function of isolating one RR resonance in each output channel. Each channel has a -3dB passband width equal to half the channel spacing. The output channel frequency spacing is equal to the FSR of the RR. The central stage has the function of ensuring to a good approximation that the centre frequency of each AWG output channel passband tracks the centre frequency of their associated RR resonance. It consists of a parallel mirrored pair of nominally identical Mach-Zehnder delay interferometers (MZDI) with FSR equal to the AWG channel spacing and hence to the FSR of the RR. The construction of the MZDI can be made either by using 2×2 directional couplers or multimode interference (MMI) couplers. The four outputs of the central stage are connected via equal optical path length waveguides to four of five AWG input channels omitting the centre channel. The AWG input channel frequency spacing is equal to half the AWG output channel spacing. The Mach-Zehnder interferometer (MZI) that precedes the MZDI pair is used as a switch that selects the active MZDI and hence the active pair of AWG input channels of the four equipped AWG input channels.

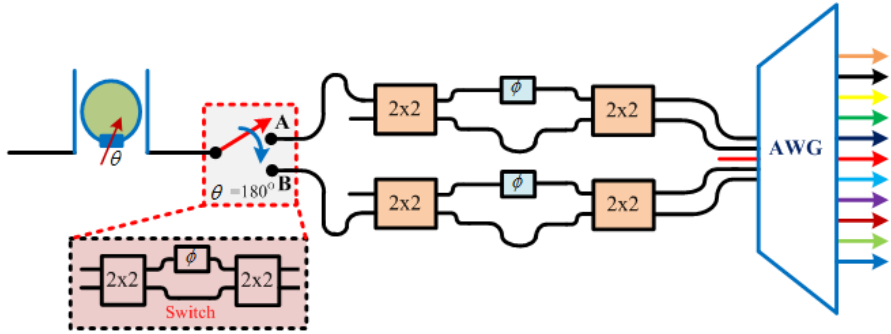

Fig. 1. Schematic of the proposed spectrometer. AWG, arrayed waveguide grating.

The spectrometer has two controls: a first control to tune the RR and a second to toggle the MZI switch. A phase change of $\Delta\theta \in [0, 2\pi]$ within the RR provided by a phase shift element is used to scan cyclically the RR resonant frequency comb over one FSR. There are two phases within this operating cycle that correspond to the state of the switch. The switch toggles at the mid-point and the end point of the scan as determined by the first control. These two controls are the only dynamic controls needed. However, it is prudent to equip the MZDI stages with quasi-static (pre-set) phase trimmers to compensate any phase bias errors due to fabrication process variations. There is some freedom of choice in the selection of the FSR and hence the total number of AWG output channels required to operate over the entire C-band. Reported experimental demonstrations of the components, our device and circuit simulations, and the process development kit support the practicality of the demonstration of a spectrometer that combines a $50\,GHz$ RR and a $88 \times 50\,GHz$ AWG.

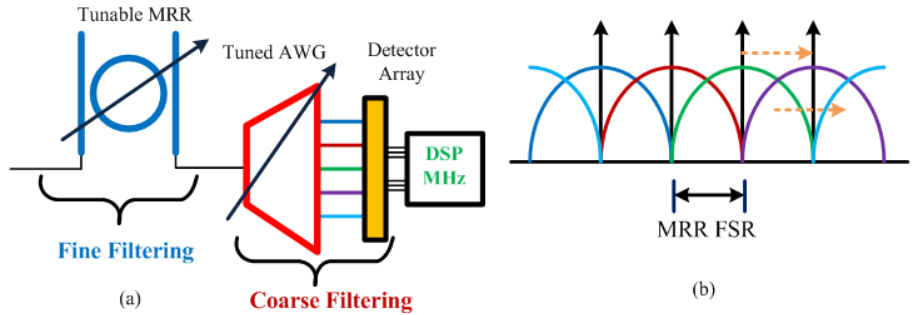

Fig. 2. (a) Tunable MRR & AWG; DSP, digital signal processing; (b) Tracking of MRR comb (grey arrows) & AWG channels (peaked curves).

To isolate an individual RR resonance within an AWG channel, ideally the comb of AWG passband centre frequencies tracks the comb of RR resonant frequencies as the spectrometer is scanned, as illustrated schematically in Fig. 2 (a, b). The benefits of ganged tuning are: minimisation of excess loss; tuning invariant channel transmission; and, an AWG adjacent channel crosstalk that is the best value (passband centre) rather than the worst value (passband edges). The digital signal processing (DSP) can be further optimized for an increased in effective measurement resolution. The AWG output channel spectra may be translated in frequency by translating the input waveguide across the input aperture of its first star coupler. An optical phased array (OPA) can be used to perform the translation of the input field profile. A necessary condition is that translation per unit frequency change of the OPA output equals that of the AWG input. The OPA may be based on a secondary AWG having an FSR equal to the primary AWG output channel frequency spacing. The secondary AWG output star coupler and the primary AWG input star coupler may be merged. However, the defocus caused by opposite sign curvature of the secondary AWG image field and the principal AWG object field strictly should be compensated by a field lens within the merged couplers. An alternative is to accept a discrete approximation to the translation of the input field profile and connect the two-star couplers by $n$ waveguides of the same optical path length. In the discrete case the output channel frequency spacing of the OPA should equal the input channel frequency spacing of the primary AWG and the OPA may be implemented by a dimension $n$ generalised Mach-Zehnder Interferometer (GMZI) rather than by a secondary AWG.

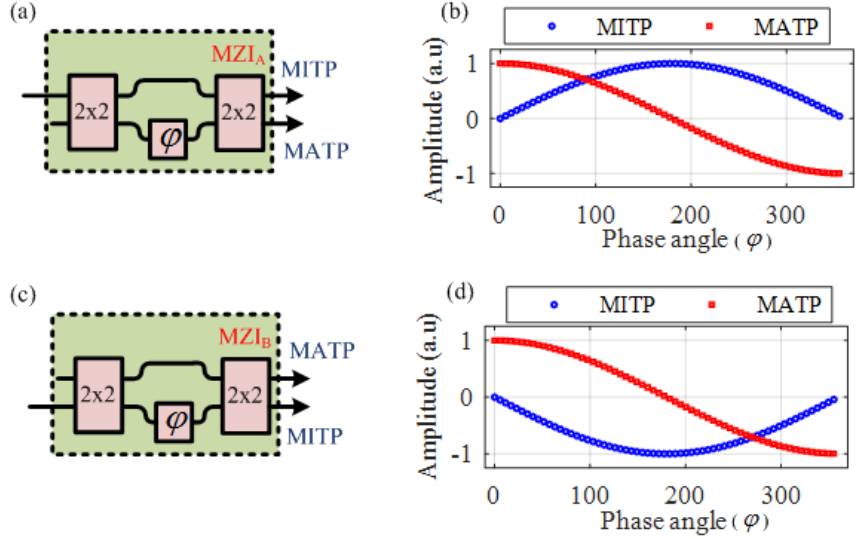

Fig. 3. Schematic of an MZI with input from upper port (a) and lower port (c); and corresponding (b) & (d) amplitude transmission plot. MITP, minimum transmission point; MATP, maximum transmission point.

A more serious problem, other than defocus, follows from field being split between the two extrema of the beam steering aperture at the frequency at which the field profile should fly-back. Over the fly back transition in the transmission this result, in two adjacent MRR resonances being passed via the same output port giving rise to unacceptable crosstalk at the edges of the tuning range. The remedy is to use two OPA-AWG units offset in frequency and select the unit offering low crosstalk at a particular tuning. With this expedient it is sufficient to use a Mach-Zehnder interferometer (MZI) as a dimension $n=2$ discrete approximation to an OPA.

Consider an MZI–AWG with input channel frequency spacing and output channel bandwidth equal to one half of its output channel frequency spacing $\Delta\omega$. A $2 \times 2$ MZI has an amplitude transmission matrix:

$$T_{MZI} = \begin{bmatrix} \sin(\varphi/2) & \cos(\varphi/2) \\ \cos(\varphi/2) & -\sin(\varphi/2) \end{bmatrix} \tag{1}$$

where $\varphi$ is the phase imbalance between its arms. Take the MZI port with $\cos(\varphi/2)$ dependence to be connected to the reference input channel of the AWG and the MZI port with $\sin(\varphi/2)$ dependence to be connected to the upshifted frequency input channel.

The amplitude transmission of a given output port is:

$$T_{AWG}(\omega) = H(\omega)\cos(\varphi/2) + H(\omega - \Delta\omega/2)\sin(\varphi/2) \tag{2}$$

where $H(\omega)$ is the transmission function of an AWG which may be taken as real to good accuracy and $\omega$ is the frequency offset from the centre of the passband. Now let us consider that $\varphi = \omega\tau$, where $\tau = 2\pi/\Delta\omega$, is the delay must be applied in one of the arms of the MZI. In this case (50 GHz output channel frequency spacing) the delay is chosen to set an FSR of 50 GHz. It follows that,

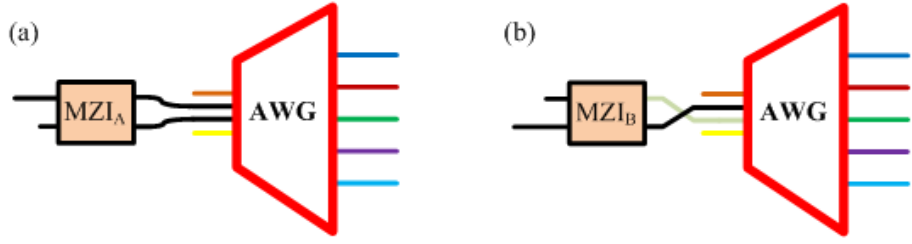

Fig. 4. MZI-AWG combination to correct the fly-back problem; (a) light is injected from the upper input port; (b) light is injected from the lower port.

$$\begin{aligned} T_{AWG}(\omega) &= 1 \quad \omega = 0 \\ T_{AWG}(\omega) &= 1 \quad \omega = \Delta\omega/4 \\ T_{AWG}(\omega) &= 1 \quad \omega = \Delta\omega/2 \end{aligned} \quad (3)$$

When $\omega\tau = 0$ ($\varphi = 0$), as per Fig. 3(b) the MATP port is active only, hence the amplitude transmission of the AWG is only $H(\omega)$. For $0° < \varphi < 180°$, the amplitude transmission of the MZI output ports have the same sign and there is constructive interference between the superimposed AWG amplitude transmissions. This corresponds to the desired frequency tracking behavior as described by Eq. (3). As the input channel frequency spacing of the AWG is set to 25 GHz, the MZI-AWG combination track over half of the output channel frequency spacing. When $180° < \varphi < 360°$, as shown in Fig. 3(b) the amplitude transmission of the MZI are opposite in sign resulting in destructive interference between the superimposed AWG amplitude transmissions. As a result, half the FSR is taken up by the undesired fly-back response.

$$\begin{aligned} T_{AWG}(\omega) &= -1 \quad \omega = -\Delta\omega/2 \\ T_{AWG}(\omega) &= 0 \quad \omega = -\Delta\omega/4 \\ T_{AWG}(\omega) &= 1 \quad \omega = 0 \end{aligned} \quad (4)$$

The sign may be corrected using an alternative MZI-AWG arrangement with the MZI port with $\cos(\varphi/2)$ dependence connected to the reference input channel of the AWG and the MZI port with $-\sin(\varphi/2)$ connected to the upshifted frequency input channel. This involves exciting the alternate input port of the MZI and connecting its output ports to the AWG inputs via a cross-over connection as shown in Fig. 4(b). The new form of the previously described amplitude transmission at the same output port becomes,

$$T_{AWG}(\omega) = H(\omega)\cos(\varphi/2) - H(\omega - \Delta\omega/2)\sin(\varphi/2) \quad (5)$$

The new frequency tracking behavior can be summarised as:

$$\begin{aligned} T_{AWG}(\omega) &= 1 \quad \omega = 0 \\ T_{AWG}(\omega) &= 0 \quad \omega = \Delta\omega/4 \\ T_{AWG}(\omega) &= -1 \quad \omega = \Delta\omega/2 \end{aligned} \quad (6)$$

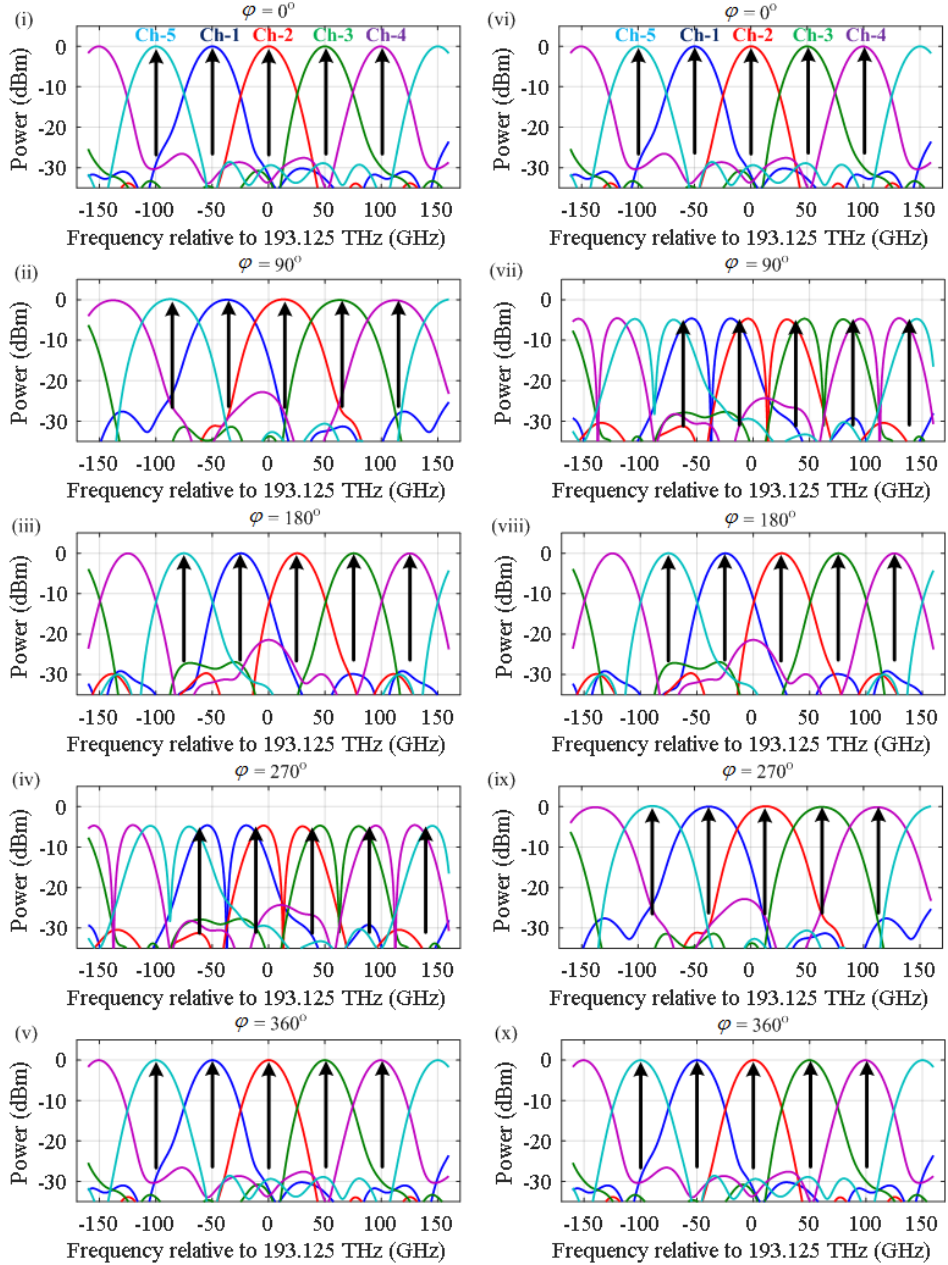

Fig 5. The amplitude transmission of an MZI-AWG combination at different phase bias of the MZI; (i)-(v) when light is launched from the upper input port of the MZI; (vi)-(x) when light is launched from the lower input port of the MZI. The black vertical arrows illustrate the frequencies of the MRR resonances associated with the phase bias.

$$T_{AWG}(\omega) = 1 \quad \omega = -\Delta\omega/2$$
$$T_{AWG}(\omega) = 1 \quad \omega = -\Delta\omega/4 \tag{7}$$
$$T_{AWG}(\omega) = 1 \quad \omega = 0$$

It shows that when $MZI_A$-AWG combination starts flying-back, $MZI_B$-AWG combination performs the tracking. Fig. 5(i-x) shows the simulation result of the tracking behavior of the circuit as shown in Fig. 4(a, b) at different MZI phase bias. As found in the earlier analysis, the MZI$_A$-AWG combination perform the frequency tracking behavior up to $\Delta\omega/2$ or 180° phase bias of the MZI, whereas MZI$_B$-AWG tracks the frequency from 180° to 360° phase bias. A phase bias ranging from 180° to 360° is equivalent to a phase bias ranging from 180° to 0°. Consequently, the circuit architectures track in opposite directions the same 0 GHz to 25 GHz span of the 50 GHz channel frequency spacing. To overcome this problem, the resolution during the fly-back phase MZI$_A$-AWG is to hand over to an alternative MZI$_B$-AWG with the MZI port with $\cos(\varphi/2)$ dependence connected to the reference input channel of the AWG and the MZI port with $-\sin(\varphi/2)$ dependence connected to the down-frequency shifted input channel of the AWG. This arrangement advantageously eliminates the cross-over interconnection. The amplitude transmission at the same output port is modified to:

$$T_{AWG}(\omega) = H(\omega)\cos(\varphi/2) - H(\omega + \Delta\omega/2)\sin(\varphi/2) \tag{8}$$

which during the fly-back phase of MZI$_A$ gives the desired frequency tracking over the -25 GHz to 0 GHz span of the 50 GHz channel frequency spacing:

$$\begin{aligned} T_{AWG}(\omega) &= 1 \quad \omega = -\Delta\omega/2 \\ T_{AWG}(\omega) &= 1 \quad \omega = -\Delta\omega/4 \\ T_{AWG}(\omega) &= 1 \quad \omega = 0 \end{aligned} \tag{9}$$

Note that at $\omega = 0, \pm\Delta\omega/2$ where handover occurs the two MZI-AWG channel outputs agree. Consequently MZI$_A$-AWG and MZI$_B$-AWG combine with handover to successfully track the full span (-25 GHz to +25 GHz) of the channel frequency spacing (50 GHz). To economise on hardware, one would like to use a switch to select the MZI input port and rearrange the connection of its two outputs between a three input AWG as shown in Fig. 6.

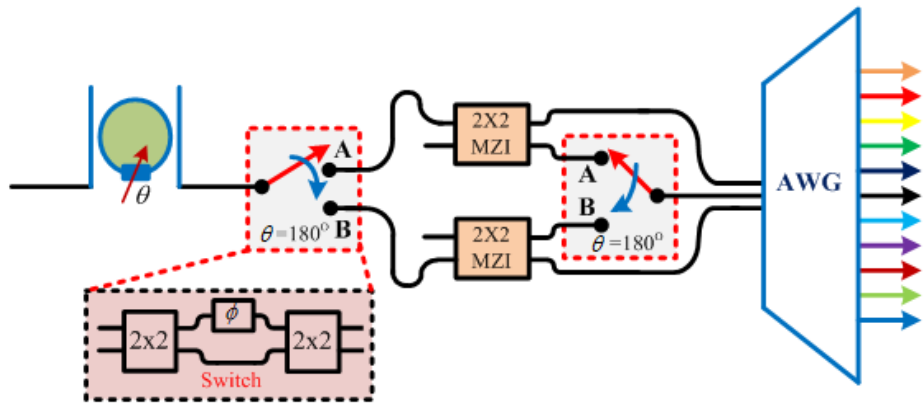

Fig. 6. Schematic diagram of the circuit architecture that can overcome the fly back problem and track the

However, a switch to rearrange the MZI outputs could introduce phase errors. It is consequently preferable to use two nominally identical MZDI to drive two pairs of adjacent ports separated by an unequipped central input port of a 5-input port AWG as shown in Fig.1. The handover is then between the set of output ports and the same set displaced by one port. The data acquisition system can resolve this shift trivially.

VPIphotonics software simulation is used to validate the theoretical prediction. Fig. 7 shows a schematic of the spectrometer stripped of the RR and with waveguide delay-lines and frequency independent static phase shift elements denoted by the symbol $\tau$. The phase shifts are adjusted manually to bias the upper and lower MZI stages and to switch between them. The switch is used to select the upper MZI in the first phase illustrated in Fig. 5(i-v) and the lower MZI in the second phase illustrated (Fig. 5(vi-x)).

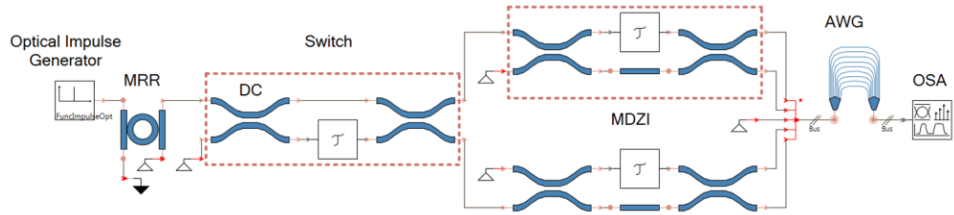

Fig. 7. VPI schematic of a circuit architecture for the purpose of demonstrating the operational principles by simulation; DC, directional coupler; OSA, optical spectrum analyzer.

To achieve auto-tuning the phase-shift elements are replaced by the delay lines. In this case (50 GHz output channel frequency spacing) the delay is chosen to set an FSR of 50 GHz. For a $Si_3N_4$ waveguide operating in the C-band the group index is almost 2 and so a delay line length of approximately 3 $mm$ is required. Fig. 8 shows the output of the AWG when both MZI are auto-tuned in this way. The flat passband and steep transitions between pass and reject bands are notable. The channel spectra corresponding to the two different switch states are interleaved

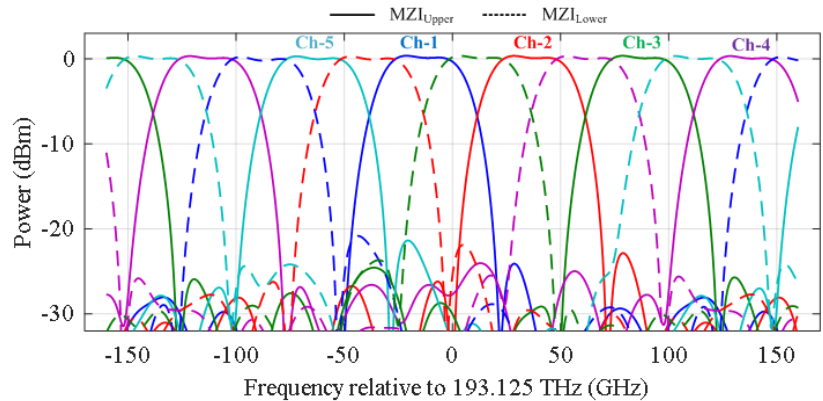

Fig. 8. AWG output channel spectra when both MZDI are auto-tuned by a delay line with an FSR equal to the AWG output channel frequency spacing. The channel spectra corresponding to the two different switch states are interleaved.

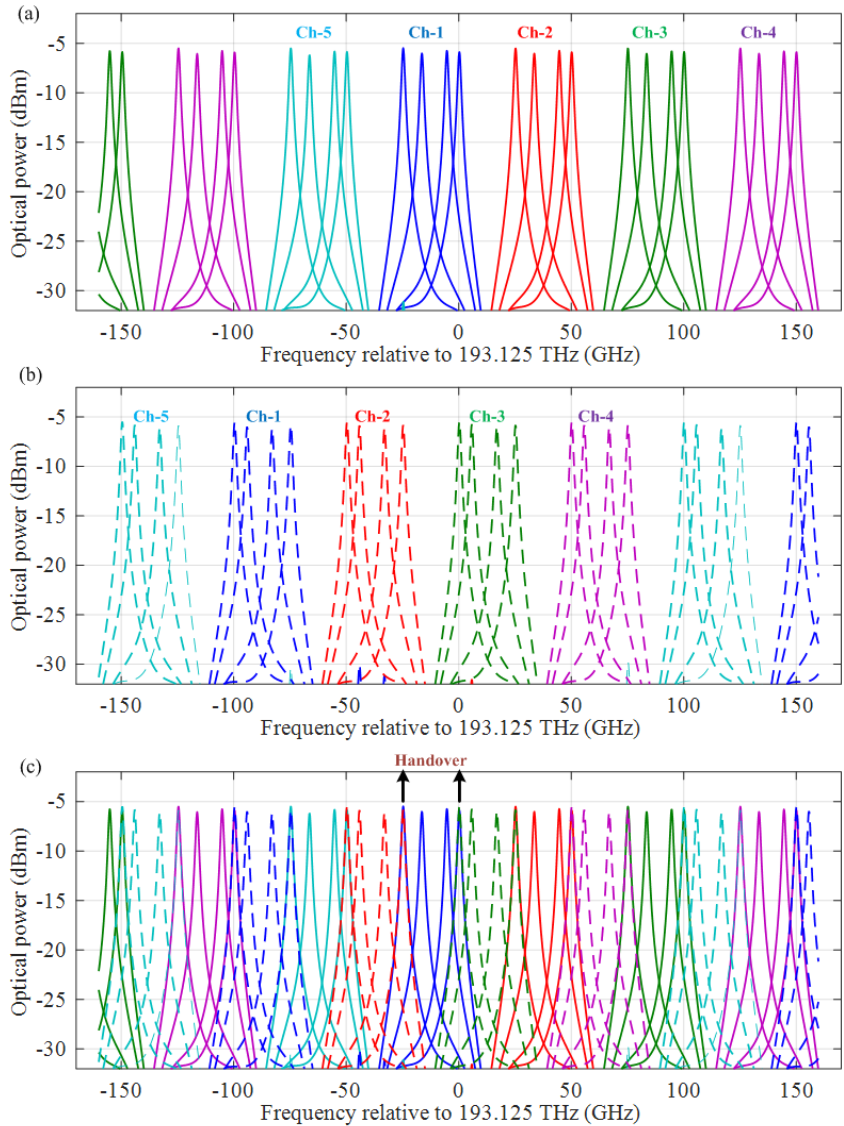

Fig. 9. A simulation of the complete spectrometer circuit with the MRR in place (as in Fig. 1) illustrating operation over complete scanning cycle over the full 50 GHz MRR FSR. (a) Upper MZDI selected phase of scanning cycle; (b) lower MZDI selected phase of scanning cycle; (c) full scanning cycle.

Fig. 9 illustrates the output channel spectra over the full scanning cycle of the complete spectrometer of Fig. 1. The switched MZDI-AWG succeeds in isolating one MR resonance with sub-GHz bandwidth within each AWG channel; the peak of the resonance is substantially constant as the spectrometer is scanned across the 50 GHz FSR of RR thereby providing a continuous spectrum across the whole operating band; and, the adjacent channel crosstalk is negligible. It may be observed that the selection of the upper MZI provides a flat response over half the channel frequency spacing corresponding to a phase bias of 0 to $-\pi$ radians and selection of the lower MZI provides a flat response over the remaining half of the channel frequency spacing corresponding to a phase bias of $-\pi$ to $-2\pi$ radians. Moreover, the readings of either selection agree at the two handover transitions in the scanning cycle one at a phase bias of $0, -2\pi$ and one at a phase bias of $-\pi$. The contiguous flat response across the band is a noticeable feature.

The spectra shown in Fig. 8 provide intuition that the spectrometer may be understood as an RR working into a pair of passband-flattened AWGs with interleaved output channel spectra. However, this does not imply that any approach to a passband flattening mechanism will suffice. The mechanism of choice is ganged tuning of the RR and AWG via (a good approximation to) translation of the input across the aperture of the first star coupler. This mechanism offers not only the best of low adjacent channel leakage found at the centre of channel passband rather than the worst of adjacent channel leakage found at the edges of the channel of the basic AWG but also it offers a means to merge the two interleaved AWGs into a single consolidated AWG which halves the number of output channels; the number of photodetectors; and the footprint size otherwise required.

To emulate the fabrication error on the circuit performance, the MZI before the AWG is configured to have less than 20 dB extinction only for the upper port and less than 30 dB for the lower port. For a stable process the same component on an integrated circuit will be almost identical. In which case, the lower output port of the MZI will always have the maximum extinction. In this example, the first directional coupler is designed to have a 56%:44% power splitting and whereas the second directional coupler has a splitting ratio of 55%:45%. It is reasonable to assume that all MZMs on the same integrated circuit have similar extinction ratio as Yamazaki et al. [16] obtained less than 0.5 dB loss variation among 4 MZMs. With the impaired extinction of the MZI (20 dB), the spectrometer performs excellently albeit the crosstalk is increased by 2-5 dB at some points. Nevertheless, a crosstalk of -20 dB or less is obtained as shown in Fig. 10.

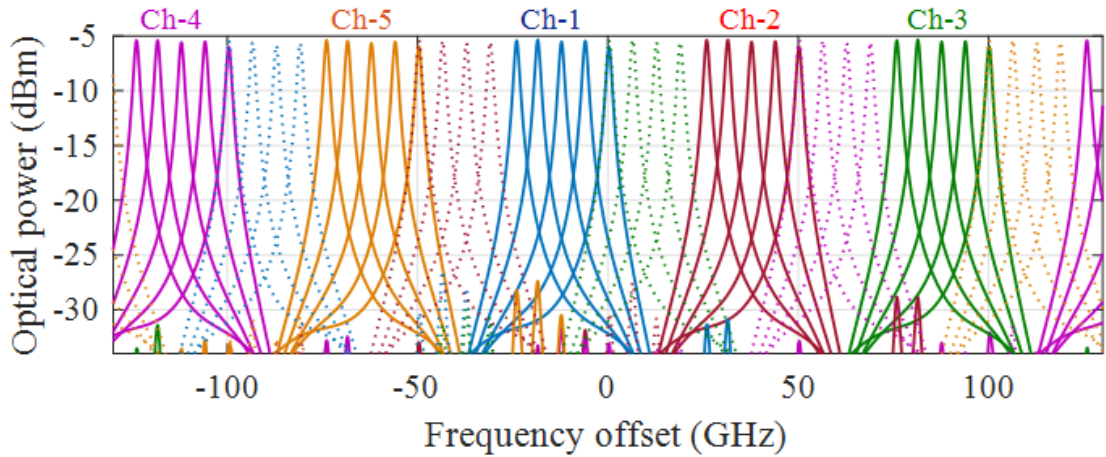

The proposed spectrometer can be fabricated in any mature fabrication platform. However, the resolution bandwidth of the spectrometer primarily depends on the insertion loss (dB/turn) of the RR along with loss in AWG. Hence in order to meet the specification of the proposed spectrometer, the CMOS compatible $Si_3N_4$ photonic integration platform has been selected as it offers the low loss, tight confinement, and low dispersion and a mature thermo-optic phase shifter technology. There are ample reports in the literature of the technological verification of SSC, MMI, tunable

RR, and sub-circuits such as MZ(D)I [11-13] and modest port count AWG fabricated using the Si$_3$N$_4$ integration platform [14]. Seyringer has reported impressive agreement of simulations and experimental measurements of a $160 \times 50\ GHz$ AWG fabricated in Si$_3$N$_4$ with footprint of ~ 1cm$^2$ [14]. Her biomedical sensing application dictates the operating center vacuum wavelength of 850 nm. However, a reference design for 850 nm may be mapped to 1550 nm by increasing all dimensions by a factor of 1.8 at least as a good starting point for further optimisation. Moreover, the spectrometer requires almost half the number of output channels which will scale down the footprint.

In summary, a state of the art on-chip spectrometer with sub-GHz resolution over the entire C-band in a compact footprint has been proposed and verified by simulation in this article. To the best of author's knowledge there have been no reports of simulation or experimental studies of a sub-GHz integrated spectrometer operating over the entire C-band and the solution proposed herein is unique. The proposed circuit architecture is feasible for photonic integration in CMOS compatible Si$_3$N$_4$ platform owing to its low loss mature technology.

*Funding*

Huawei Canada sponsored research agreement 'Research on ultra-high resolution on-chip spectrometer'.

*Acknowledgments*

Mehedi Hasan acknowledges the Natural Sciences and Engineering Research Council of Canada (NSERC) for their support through the Vanier Canada Graduate Scholarship program. Trevor J. Hall is grateful to HUAWEI, Canada for their support of this work. Trevor J. Hall is also grateful to the University of Ottawa for their support of a University Research Chair.

*Disclosures*

The authors declare no conflicts of interest.